\documentclass[10pt,letterpaper]{IEEEtran}
\usepackage[utf8]{inputenc}
\usepackage[T1]{fontenc}
\usepackage{graphicx}
\usepackage{grffile}
\usepackage{longtable}
\usepackage{wrapfig}
\usepackage{rotating}
\usepackage[normalem]{ulem}
\usepackage{amsmath}
\usepackage{amssymb}
\usepackage{amsfonts}
\usepackage{theorem}
\usepackage{bm}
\usepackage{textcomp}
\usepackage{capt-of}
\usepackage{hyperref}
\usepackage{cite}
\usepackage{booktabs}
\usepackage{blindtext}

\newcommand{\letterref}[2]{\hyperref[#1]{\ref*{#1}#2}}

\usepackage{mdframed}
\newtheorem{problem}{Problem}
\newtheorem{Prop}{Proposition}
\newtheorem{Theorem}{Theorem}

\usepackage{tikz}
\usetikzlibrary{positioning}
\tikzset{>=latex}
\usepackage{pgfplots}
\usepgfplotslibrary{fillbetween}
\usepackage{pgfplotstable}

\DeclareMathOperator{\Ima}{Im}

\long\def\beginpgfgraphicnamed#1#2\endpgfgraphicnamed{\includegraphics{#1}}

\usepackage{shortcuts_OPT}

\author{
\IEEEauthorblockN{T. Mitchell Roddenberry and Santiago Segarra}
\thanks{Research was sponsored by the Army Research Office and was accomplished under Cooperative Agreement Number W911NF-19-2-0269. The views and conclusions contained in this document are those of the authors and should not be interpreted as representing the official policies, either expressed or implied, of the Army Research Office or the U.S. Government. The U.S. Government is authorized to reproduce and distribute reprints for Government purposes notwithstanding any copyright notation herein. Both authors are with the Department of Electrical and Computer Engineering, Rice University. Emails: \url{mitch@rice.edu}, \url{segarra@rice.edu}}
}
\title{HodgeNet: Graph Neural Networks for Edge Data}
\hypersetup{
 pdfauthor={T. Mitchell Roddenberry and Santiago Segarra},
 pdftitle={HodgeNet: Graph Neural Networks for Edge Data},
 pdfkeywords={},
 pdfsubject={},
 pdflang={English}}
\begin{document}

\maketitle

\begin{abstract}
  Networks and network processes have emerged as powerful tools for modeling social interactions, disease propagation, and a variety of additional dynamics driven by relational structures.
  Recently, neural networks have been generalized to process data on graphs, thus being able to learn from the aforementioned network processes achieving cutting-edge performance in traditional tasks such as node classification and link prediction.
  However, these methods have all been formulated in a way suited only to data on the nodes of a graph.
  The application of these techniques to data supported on the edges of a graph, namely \textit{flow signals}, has not been explored in detail.
  To bridge this gap, we propose the use of the so-called \textit{Hodge Laplacian} combined with graph neural network architectures for the analysis of flow data.
  Specifically, we apply two graph neural network architectures to solve the problems of flow interpolation and source localization.
\end{abstract}

\section{Introduction}
\label{sec:intro}

As relational, structured data becomes increasingly ubiquitous, network science and graph analytics grow in importance for modeling data in many domains~\cite{Newman2010,Strogatz2001,Jackson2010, segarra_2015_authorship, medaglia_2017_brain}.
Recently, deep learning techniques have been generalized to data supported on the nodes of graphs, enabling cutting-edge results in node classification, graph classification, and link prediction~\cite{Bronstein2017}.
The main theme of these approaches is the application of a graph shift operator that generalizes the time delay or shift operator in classical signal processing.
By learning polynomials of the graph shift operator coupled with nonlinear activation functions, neural networks can be generalized to the graph domain.

In many relevant settings, the data of interest is supported most naturally on the edges of a graph, such as a \textit{flow} -- modeling the transfer of mass, energy, or information -- through a network.
Recently, Hodge theory has been applied to higher-order graph structures (represented by simplicial complexes) on which such data is supported.
In this context, the spectral properties of the \textit{Hodge Laplacian} have been leveraged to solve the problems of flow denoising~\cite{SchaubDenoising}, flow interpolation~\cite{JiaFlowSSL}, and higher-order network topology inference~\cite{BarbarossaTopological}.
In the current paper, we propose to combine these advances in graph signal processing (GSP) for flow data along with the recent successes of graph neural networks (GNNs) to solve inverse problems on graph flow data, specifically flow interpolation and source localization.

\vspace{1mm}
\noindent
{\bf Related work.}
Several problems related to signals defined on higher-order graph structures have recently been considered.
Flow denoising through low-pass graph filters based on the Hodge Laplacian has been studied in~\cite{SchaubDenoising}.
Furthermore, \cite{JiaFlowSSL}~proposed interpolation and sampling methods for flow data under priors on the signal characteristics with respect to the Hodge Laplacian, addressing the same flow interpolation problem discussed in Section~\ref{sec:problem-interpolation}.
Most recently, \cite{BarbarossaTopological} and \cite{BarbarossaLearning}~employed these tools to tackle the problem of topology inference from data on abstract simplicial complexes.
Outside of the context of signal processing, the Hodge Laplacian for graphs has enabled novel data science approaches. 
These include a spectral method for pairwise ranking in sports analytics~\cite{Jiang2011} and the formal characterization of random walks in (higher-order) simplicial complexes~\cite{SchaubRandom}.

\vspace{1mm}
\noindent
{\bf Contributions.}
The primary contribution of this paper is the first integration of discrete Hodge theory with deep learning techniques by incorporating the Hodge Laplacian in GNN architectures.
Additionally, our methods provide a prior-free approach to the problem of flow interpolation, in contrast with~\cite{JiaFlowSSL}.
Finally, we use the aggregation GNN architecture introduced by~\cite{GamaConvolutional} to solve a source localization problem for flow data.

\section{Background}
\label{sec:bg}

\subsection{Graphs, graph signals, and graph signal processing}
\label{sec:bg-gsp}

A \textit{graph} is a data structure consisting of a set of nodes ${\cal V}$ connected by a set of edges ${\cal E}\subseteq{\cal V}\times{\cal V}$, denoted by ${\cal G}=(\cal{V,E})$.
An \textit{undirected graph} has an edge set consisting of unordered tuples, i.e., $(i,j)\in{\cal E}\iff(j,i)\in{\cal E}$.
For convenience, we will indicate the cardinality of the node and edge sets as $N:=|{\cal V}|$ and $E:=|{\cal E}|$, respectively.
Graphs can be compactly represented by their \textit{adjacency matrix} ${\bm A}\in{\mathbb R}^{N\times N}$, where for some indexing of ${\cal V}$ with the integers $\{1,2,\ldots,N\}$, we have ${\bm A}_{ij}=1$ if $(i,j)\in{\cal E}$ and ${\bm A}_{ij}=0$, otherwise. An alternative representation is the \textit{incidence matrix} ${\bm B}\in{\mathbb R}^{N\times E}$, where for the same indexing of ${\cal V}$ and a labeling of ${\cal E}$ with $\{e_i\}_{i=1}^E$ we have
\begin{equation}
  \label{eq:incidence}
  {\bm B}_{ij}=
  \begin{cases}
    -1 & e_j=(i,k)\text{ for some }k\in{\cal V}, \\
    1 & e_j=(k,i)\text{ for some }k\in{\cal V}, \\
    0 & \text{ otherwise}.
  \end{cases}
\end{equation}
Notice that this assigns an inherent direction or orientation to each edge $e_j$.
For undirected graphs, this choice is arbitrary but does not affect the subsequent definitions of other relevant graph matrices in a meaningful way.

While the inherent structure of a graph is often of interest, GSP offers tools for the analysis of signals supported by graphs~\cite{EmergingFieldGSP, ortega_2018_graph}.
In the same way that a discrete-time signal is a function from the integers to the real numbers, a \textit{graph signal} maps nodes to real numbers, $x\colon{\cal V}\to{\mathbb R}$.
For some node labeling with the integers $\{1,2,\ldots,N\}$, a graph signal $x$ can be conveniently represented as a vector ${\bm x}\in{\mathbb R}^N$.
Extending the analogy with discrete-time signal processing, a \textit{graph filter} $g$ is defined as a polynomial of a graph shift operator ${\bm S}$~\cite{segarra_2017_optimal}
\begin{equation}
  \label{eq:graphfilter}
  g\star {\bm x} = \sum_{k=0}^{K-1}g(k){\bm S}^k{\bm x}.
\end{equation}

A common choice of graph shift operator is the \textit{graph Laplacian}, defined as
\begin{equation}
  \label{eq:graphlaplacian}
  {\bm L}_0={\bm D}-{\bm A},
\end{equation}
where ${\bm D} = \diag(\bm A \mathbf{1})$ is the diagonal matrix of node degrees.
The eigenvectors of the graph Laplacian (ordered in terms of increasing associated eigenvalues) provide an orthonormal basis of frequencies that is specific to the graph at hand. 
Consequently, the coefficients of a signal projected onto the eigenbasis of the graph Laplacian can be interpreted as a measurement of the smoothness of said signal in terms of local uniformity.
We refer to~\cite{EmergingFieldGSP, ortega_2018_graph} for a more complete discussion of GSP basics.

\subsection{Graph neural networks}
\label{sec:bg-gcn}

GNNs use tools from GSP combined with traditional convolutional neural network architectures to process data on graphs.
Convolutional neural networks for images are typically composed of many layers, each consisting of a localized filtering operation followed by a nonlinear activation function~\cite{DeepLearningBook}.
As laid out in~\cite{DefferardConvolutional} and \cite{kipf2017semi}, these layers can be generalized to graphs simply by replacing the convolution operation with a graph filter as defined in \eqref{eq:graphfilter}.
Moreover, one can restrict the order $K$ of the filter to constrain the filter to operate locally on each node.
By then applying a nonlinear function, we form a graph convolutional layer.
Architectures following this form have achieved state of the art results in tasks such as node classification, graph classification, and link prediction~\cite{DefferardConvolutional,kipf2017semi}.
We defer discussions of our proposed architectures to Section~\ref{sec:hn}.

% \begin{figure}
%   \centering
%   \resizebox{0.3\linewidth}{!}{\input{gradient}}
%   \qquad
%   \resizebox{0.3\linewidth}{!}{\input{cyclic}}
%   \caption{Examples of gradient and cyclic flows. (A) A purely gradient flow. The potentials $\Phi$ at each node are indicated by their color, and the induced flow signal is indicated by edge thickness and direction. (B) A purely cyclic flow, where the net flow incident to each node is zero.}
%   \label{fig:decomp}
% \end{figure}

\subsection{Flow signals and the Hodge Laplacian}
\label{sec:bg-flow}

So far, our discussion has considered signals on the nodes of a graph, i.e., $x\colon{\cal V}\to{\mathbb R}$.
We define \textit{flow signals} as signals on the edges of a graph $f\colon{\cal E}\to{\mathbb R}$ that have the property of skew-symmetry:
\begin{equation}
  \label{eq:skewsym}
  f(i,j)=-f(j,i), \,\,\, \forall(i,j)\in{\cal E}.
\end{equation}
Similar to graph signals, for some edge labeling with the integers $\{1,2,\ldots,E\}$ and a choice of edge orientations, a flow signal $f$ can be represented as a vector ${\bm f}\in{\mathbb R}^E$.
It is important to note that the following discussion does not depend on the chosen edge labeling and orientations.

As pointed out by~\cite{SchaubDenoising}, a tempting shift operator for flow signals is the \textit{linegraph Laplacian}, formed by treating each edge in the original graph as a `node' and treating incidence between edges in the original graph as (unweighted) `edges.'
One would then consider the graph Laplacian matrix of this construction, which we refer to as ${\bm L}_\mathrm{LG}$, and apply standard GSP methods.
Similar to the graph Laplacian, the spectral characteristics of ${\bm L}_\mathrm{LG}$ relate to the local uniformity (smoothness) of flow signals.

An alternative graph shift operator for flow signals is the \textit{Hodge Laplacian}, defined in terms of the graph incidence matrix ${\bm B}$ [cf.~\eqref{eq:incidence}]
\begin{equation}
  \label{eq:hodgelaplacian}
  {\bm L}_1={\bm B}^\top{\bm B}.
\end{equation}
In contrast with the smoothness measured by ${\bm L}_0$ and ${\bm L}_{\mathrm{LG}}$, the Hodge Laplacian captures \textit{conservatism} of flow signals~\cite{BarbarossaTopological}.
Conservative flows are referred to as \textit{cyclic} since they are naturally composed of cycles, or loops, in the graph.
Non-conservative flows are referred to as \textit{gradient} since they are induced by differences in potentials on the nodes of the graph.
That is, for a purely gradient flow ${\bm f}$, there is a vector of node potentials ${\bm \Phi}$ such that ${\bm f}={\bm B}^\top{\bm \Phi}$.
This distinction is formally captured in the Hodge decomposition for graphs~\cite{SchaubDenoising, BarbarossaTopological}.
\begin{Theorem}\label{thm:helmholtz}
  The space of edge-flows admits an orthogonal decomposition into a cyclic component and a gradient component
  \begin{equation}\label{eq:helmholtz}
    \mathbb{R}^E=\ker\left({\bm L}_1\right)\oplus\Ima\left({\bm B}^\top\right).
  \end{equation}
\end{Theorem}
In~\eqref{eq:helmholtz}, $\ker\left({\bm L_1}\right)$ is a basis for the space of cyclic edge flows, and $\Ima\left({\bm B}^\top\right)$ is a basis for the space of gradient edge flows.
A more general version of this decomposition exists for flow signals on higher-order simplicial complexes, but we focus on the case for simple graphs here.

\section{Problem statement}
\label{sec:problem}

To motivate our proposed methods, we introduce two problems in the analysis of flow data on graphs.
The first is a regression problem for the purpose of flow interpolation, while the second is a graph-level signal classification problem.

\subsection{Flow interpolation}
\label{sec:problem-interpolation}

A common task in image processing is \textit{inpainting}, where some pixels of an image are missing, and we wish to use the surrounding image information to predict their values.
We specify an analogous problem for flow signals next.
\begin{problem}[Flow interpolation]\label{prob:interpolation}
  Given a graph ${\mathcal G}=({\mathcal V},{\mathcal E})$ and an observation of a flow signal $f:{\mathcal E}\to{\mathbb R}$ over a subset $\Omega\subseteq{\mathcal E}$, infer the value of $f$ over the unobserved edges ${\mathcal E}\setminus\Omega$.
\end{problem}
For convenience, we indicate the flow over the observed set as $f_\Omega$, and the flow over the unobserved set as $f_{{\mathcal E}\setminus\Omega}$.
We refer to the inferred flow as $\widehat{f}$.

In the context of GSP, Problem~\ref{prob:interpolation} has already been addressed in~\cite{JiaFlowSSL}, where a flow-conservation prior is placed on $f$, and $\widehat{f}$ is recovered via a convex optimization procedure that minimizes the gradient component of the recovered flow.
Our aim in applying GNNs to this problem is to remove the dependence on a prior, alleviating problems that arise when the model does not match the true flow, as illustrated in Section~\ref{sec:exp-interp}.

\subsection{Source localization}
\label{sec:problem-localization}

In many scenarios, flow on a graph is mainly induced by one or several sources.
For instance, in an urban setting, the flow of traffic in the afternoon would mostly be induced by the exodus of cars from the city center to surrounding neighborhoods.
In the source localization task, we aim to recover the source of an observed flow signal.
\begin{problem}[Source localization]\label{prob:localize}
  Given a graph ${\mathcal G}=({\mathcal V},{\mathcal E})$ with a known partition of ${\mathcal V}$ into communities ${\mathcal C}_1,{\mathcal C}_2,\ldots,{\mathcal C}_k$, as well as a flow signal $f$ induced by an unknown source node $v\in{\mathcal V}$, recover the community $1\leq j\leq k$ in which $v$ is contained, i.e., $\{ j \, | \, v\in{\mathcal C}_j \}$.
\end{problem}
Problem~\ref{prob:localize} is a simpler version of the typical source localization problem where one aims to recover the node that a graph signal originated from.
In this coarser alternative, we seek to infer the community in which the origin node is located.

\begin{figure*}
  \centering
  \includegraphics[width=0.9\linewidth]{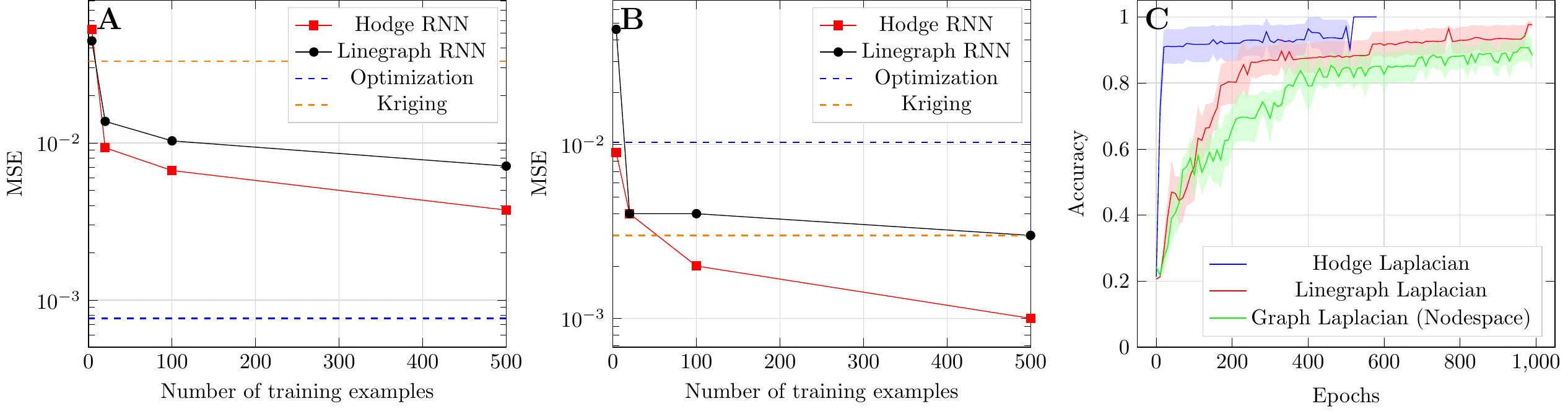}
  \caption{Performance of RNN architectures in flow interpolation task when trained on differently-sized synthetic sets of flow signals compared to kriging and optimization procedures for (A) conservative flows and (B) smooth gradient flows. (C) Convergence of test accuracy of aggregation GNN in source localization task for Hodge, linegraph, and nodespace shift operators. The shaded regions indicate the estimated standard deviation in performance over 10 training runs.}
  \label{fig:exp}
\end{figure*}

\section{HodgeNet}
\label{sec:hn}

For solving the flow interpolation and source localization problems, we propose to apply GNN architectures using the Hodge Laplacian as a shift operator.
Although the following architectures are distinct from one another, we loosely refer to them under the umbrella term \emph{HodgeNet} to indicate the use of the Hodge Laplacian as a constituent element of the neural networks.

\subsection{Hodge RNN for flow interpolation}
\label{sec:hn-rnn}

Inspired by the GNN introduced in~\cite{scarselli2009graph}, we propose a recurrent neural architecture that receives a sequence of aggregated (partially observed) flow signals as input and outputs a prediction of the flow at the unobserved edges.
This sequence is formed by repeatedly applying the Hodge Laplacian to the masked flow signal, which updates a hidden state at each step, from which the output is computed.
More precisely, the input ${\bm x}_k \in{\mathbb R}^E$, hidden state ${\bm H}_k\in{\mathbb R}^{E\times F}$ ($F$ is a parameter indicating the dimension of the feature space), and output ${\bm o}_k\in{\mathbb R}^E$ at the $k$th step are given by
\begin{align}
  {\bm x}_k &= \frac{{\bm L}_1}{\lambda_1}{\bm x}_{k-1}, \label{eq:rnnlayers1} \\
  {\bm H}_k &= \sigma\left({\bm x}_k{\bm u}^\top+{\bm H}_{k-1}{\bm V}\right), \label{eq:rnnlayers2}  \\
  {\bm o}_k &= \sigma\left({\bm H}_k{\bm w}\right), \label{eq:rnnlayers3} 
\end{align}
for trainable weights ${\bm u},{\bm w}\in{\mathbb R}^{F},{\bm V}\in{\mathbb R}^{F\times F}$, and nonlinear function $\sigma(\cdot)$ applied elementwise.
We initialize the hidden state as ${\bm H}_0={\bm 0}$ and the initial input ${\bm x}_0={\bm f}_\Omega$ is the masked flow signal with value $0$ in the unobserved entries.
In \eqref{eq:rnnlayers1}, the Hodge Laplacian is normalized by its maximum eigenvalue $\lambda_1$.
We terminate this recursive process after some pre-specified number of layers $K$.

Recalling the skew-symmetry property of flow signals \eqref{eq:skewsym}, we wish to maintain this architecture's equivariance to the arbitrary choice of edge orientations.
That is, if we denote the output of this architecture by $\bm o_K({\bm f},{\bm L}_1)$, for some input flow ${\bm f}$ with associated Hodge Laplacian ${\bm L}_1$, a diagonal ``edge-flipping'' matrix ${\bm F}\in{\mathbb R}^{E\times E}$ with diagonal entries in $\{-1,1\}$ should obey
\begin{equation}
  \label{eq:sign-equivariance}
  {\bm o}_K\left({\bm F}{\bm f},{\bm F}{\bm L}_1{\bm F}\right) = \bm{Fo}_K\left({\bm f},{\bm L}_1\right).
\end{equation}
That is, if the orientation for an edge is flipped, the output of the Hodge RNN using the reoriented Hodge Laplacian should take the same value at that edge, except with opposite sign, reflecting the skew-symmetry property of flows~[cf.~\eqref{eq:skewsym}].
This is characterized by the following proposition.
\begin{Prop}[Equivariance of Hodge RNN]\label{prop:sign-equivariance}
  The output ${\bm o}_K$ of the Hodge RNN described by \eqref{eq:rnnlayers1}-\eqref{eq:rnnlayers3} is equivariant to the arbitrary choice of edge orientations as in \eqref{eq:sign-equivariance} if and only if $\sigma(\cdot)$ is an odd function.
\end{Prop}
See Appendix~\ref{app:prop1} for the proof.
Guided by Proposition~\ref{prop:sign-equivariance}, we select $\sigma$ in \eqref{eq:rnnlayers2} and \eqref{eq:rnnlayers3} to be the (elementwise) soft-thresholding operator with a trainable threshold $\tau$, i.e., $\sigma(x; \tau) = \mathrm{sign}(x)[|x|-\tau]_+$. 
In this way, $\sigma$ is odd while maintaining an evident functional similarity to the popular ReLU nonlinearity.

For the purpose of flow interpolation, the RNN architecture is trained by artificially masking known flow signals and minimizing the mean-squared error between the ground truth and output flow over the masked edges.
That is, for some observation set $\Omega$ and a set of artificially masked edges $\Psi\subset\Omega$, we aim to find parameters ${\bm u},{\bm V},{\bm w}$ that minimize ${\mathcal L}$, defined as:
\begin{equation}
  \label{eq:rnnloss}
  {\mathcal L}=\frac{1}{\left|\Psi\right|}\left\|\left[{\bm f}-{\bm o}_K({\bm f}_{\Omega\setminus\Psi},{\bm L}_1)\right]_{\Psi}\right\|_2^2,
\end{equation}
where we use ${\bm x}_S\in{\mathbb R}^{|S|}$ to refer to the vector of elements of ${\bm x}$ corresponding to the set $S$.
This procedure can be applied both to a large set of historical flow signals or to a single partially observed flow signal by training on multiple artificial masks $\Psi$ applied to the observed edges.

\subsection{Hodge aggregation GNN for source localization}
\label{sec:hn-agg}

For the purpose of graph-level signal classification, we adapt the aggregation GNN (AGNN) proposed by~\cite{GamaConvolutional}.
Similar to the RNN approach, we repeatedly apply the Hodge Laplacian normalized by its largest eigenvalue ${\bm L}_1 / \lambda_1$ to a signal ${\bm f}$, yielding a real-valued sequence observed at a fixed set of edges in the graph.
More specifically, for $K$ observed edges indicated by a fixed binary row-selection matrix ${\bm C}\in\{0,1\}^{K\times E}$, the aggregation sampling method yields the multi-channel sequence ${\bm G}\in{\mathbb R}^{K\times E}$ given by
\begin{equation}
  \label{eq:aggsample}
  {\bm G}={\bm C}\left[{\bm f},\left(\frac{{\bm L}_1}{\lambda_1}\right){\bm f},\left(\frac{{\bm L}_1}{\lambda_1}\right)^2{\bm f},\ldots,\left(\frac{{\bm L}_1}{\lambda_1}\right)^{E-1}{\bm f}\right].
\end{equation}
We then apply a standard 1D-CNN with ReLU nonlinearities to the sequence ${\bm G}$~\cite{DeepLearningBook}. 
Once the sequence has been reduced to a sufficiently low dimension, it is fed into a fully-connected layer followed by a softmax operation for classification.
We denominate the output of this architecture $\mathrm{AGG}_\theta({\bm f},{\bm L_1})$, for input flow ${\bm f}$ with associated Hodge Laplacian ${\bm L}_1$, and CNN filter parameters collected in $\theta$.

An analogous version of Proposition~\ref{prop:sign-equivariance} for the Hodge AGNN with ReLU nonlinearities can be stated as follows.
\begin{Prop}[Invariance of Hodge AGNN]\label{prop:agg-invariance}
  For a given set of parameters $\theta$, the output of the Hodge AGNN satisfies the following:\\
  i) $\mathrm{AGG}_\theta$ is invariant to the arbitrary choice of edge orientations outside of the edges selected by ${\bm C}$, i.e., $\mathrm{AGG}_\theta({\bm F}{\bm f},{\bm F}{\bm L}_1{\bm F})=\mathrm{AGG}_\theta({\bm f},{\bm L}_1)$ if ${\bm F}_{ii}=1$ for all edges $i$ selected by ${\bm C}$. \\
  ii) For an arbitrary set of CNN parameters $\theta$ and edge-flipping matrix ${\bm F}$, there exists a rotated set of CNN parameters $\theta'$ such that $\mathrm{AGG}_\theta({\bm f},{\bm L}_1)=\mathrm{AGG}_{\theta'}({\bm F}{\bm f},{\bm F}{\bm L}_1{\bm F})$ for all ${\bm f}\in{\mathbb R}^E$.
\end{Prop}
See Appendix~\ref{app:prop2} for the proof.
The first part of Proposition~\ref{prop:agg-invariance} reveals that even if \emph{after training} we flip the orientations of the edges not selected by $\bm C$ then the output of the Hodge AGNN remains invariant. 
More importantly, the second part of the proposition states that the ultimate performance of the neural network does not depend on the edge orientations chosen \emph{before training}, if the parameters $\theta$ are randomly initialized according to a spherically symmetric distribution, \eg an i.i.d. multivariate gaussian distribution.
Notice that this result is expected and appealing, since the orientation of the edges is arbitrary and an architecture whose performance depends on the choice of these orientations would be undesirable in practice. 

\section{Numerical experiments}
\label{sec:exp}

In this section, we demonstrate the proposed methods as applied to the flow interpolation and source localization problems\footnote{Code for these experiments can be found at \url{github.com/tmrod/hodgenet}}.
To highlight the superior modeling capabilities of the Hodge Laplacian for flow data, we compare the performance of our proposed architectures to similar models using standard graph shift operators, such as the linegraph Laplacian.
That is, we replace ${\bm L}_1$ with ${\bm L}_\mathrm{LG}$ in~\eqref{eq:rnnlayers1}-\eqref{eq:rnnlayers3}~and~\eqref{eq:aggsample}.

\subsection{Flow interpolation}
\label{sec:exp-interp}

\begin{table}
  \centering
  \caption{Flow interpolation performance for the RNN architectures, the convex optimization procedure, and spatial kriging.}
  \label{tab:exp-interp}
  \begin{tabular}{lc}
    \toprule
    Method & PSNR (dB) \\
    \midrule
    Hodge RNN & 20.5 \\
    Linegraph RNN & 18.7 \\
    ConvOpt & 31.0 \\
    Kriging & 14.8 \\
    \bottomrule
  \end{tabular}
\end{table}

For the flow interpolation task, we apply the RNN architecture to the Anaheim traffic flow dataset~\cite{stabler2019transportation}.
The flow signal is artificially masked to make 10\% of edges unobserved, and the rest of the edges in the graph are observed, i.e., $|\Omega|=0.9E$ [cf.~Problem~\ref{prob:interpolation}].

The RNN architecture is applied using both the Hodge Laplacian as a shift operator and the linegraph Laplacian, with the training performed on the observed set of edges as described in Section~\ref{sec:hn-rnn}.
Since the linegraph Laplacian does not encode the orientation of flow signals, we apply it to the absolute value of the flow signal, i.e., ${\bm f}$ is replaced by $|{\bm f}|$.
These are compared to the convex optimization approach proposed by~\cite{JiaFlowSSL}, as well as a spatial kriging approach similar to~\cite{wang2009forecasting}, where the graph is embedded in space using spectral drawing~\cite{SpectralDrawing}, and the flow signals are interpolated via a Gaussian process regression.
Again, this does not capture flow orientation, so we apply it to the absolute value of the flow signal.

The results are shown in Table~\ref{tab:exp-interp}.
Clearly, the convex optimization approach outperforms all methods, which is due to the flow-conservation prior enforced by the method holding strongly.
However, among the prior-free approaches, the Hodge RNN model performs the best.
This can be attributed to the spectral properties of the Hodge Laplacian matching intuitive properties of traffic flows (i.e., conservatism), as opposed to the spectral properties of the linegraph Laplacian, which capture smoothness in the sense of uniformity.

In the previous setting, we only had access to a single partially observed flow signal over a graph.
To understand how the RNN for flow interpolation can improve with access to historical flow data, we generate a synthetic dataset by adding cyclic Gaussian noise (that is, a Gaussian random vector in ${\mathbb R}^E$ that has been projected into $\ker({\bm L}_1)$) to the original Anaheim flow signal, as well as a small amount of smooth gradient noise.
This procedure is repeated to yield multiple flow signals and the RNN architectures are trained on these datasets.
The convex optimization and kriging baselines do not learn from multiple flow signals, but we also evaluate their performance on this data for the sake of comparison. The results of this experiment for increasing training set size is shown in Figure~\letterref{fig:exp}{A}.
The behavior is as expected: The optimization approach has a very accurate prior, and thus outperforms the prior-free approaches.
Additionally, both RNN architectures improve with an increasing amount of training data, with the Hodge RNN consistently outperforming the model based on the linegraph Laplacian ${\bm L}_\mathrm{LG}$.

To emphasize the versatility of the RNN approaches (due to the lack of a strong prior), we generate a set of synthetic (smooth) gradient flows using a procedure similar to the one previously described.
In this case, the convex optimization procedure experiences a model mismatch since the flow-conservation prior does not hold.
This is illustrated in Figure~\letterref{fig:exp}{B}, where the optimization approach performs poorly and both RNN approaches improve with more historical data.
Interestingly, kriging performs comparatively better in this setting, which could be explained by a smooth gradient flow corresponding to a set of smooth node potentials~\cite{BarbarossaTopological}.

\subsection{Source localization}
\label{sec:exp-source}

For the source localization problem, we consider flows induced by a smoothly varying vector of node potentials, which is concentrated in a localized region of the graph.
To model this, the set of node potentials is obtained by applying a diffusion graph process ${\mathcal D}({\bm A})$ to an impulse signal at some \emph{source} node $v$. These node potentials then induce a flow via the incidence matrix ${\bm B}$.
More precisely, the observed flow is modeled by
\begin{equation}
  \label{eq:exp-source-model}
  {\bm f}={\bm B}^\top{\mathcal D}({\bm A})\delta_v+\epsilon,
\end{equation}
where $\delta_v$ is a graph signal taking value $1$ at node $v$ and $0$ elsewhere, and $\epsilon$ is normally distributed additive noise.
Notice that this signal model is purely gradient, aside from the added noise.

Following the experiments in~\cite{GamaConvolutional}, we consider a graph drawn from a planted partition model with $p=0.8$, $q=0.2$, and $k=5$ communities of $20$ nodes each.
A set of 10000 training signals is then generated by randomly choosing a diffusion time $1\leq t_i\leq 20$ uniformly at random for each signal indexed by $i\in\{1,2,\ldots,10000\}$, then applying the diffusion operator ${\mathcal D}_{t_i}({\bm A})=\left({{\bm A}}/{\lambda_1}\right)^{t_i}$ to an impulse $\delta_{v_i}$ as in \eqref{eq:exp-source-model}, where $v_i$ is randomly chosen among the nodes of highest degree in each community.
Here, $\lambda_1$ denotes the largest eigenvalue of $\bm A$.
The flow signal ${\bm f}_i$ is then generated as in~\eqref{eq:exp-source-model}.
The signal is then categorically labeled according to the community from which the impulse originated, e.g., if $v_i\in{\mathcal C}_j$, then the label for ${\bm f}_i$ is $j$ [cf.~Problem~\ref{prob:localize}].

An AGNN is then trained on these signals, and tested on a set of 2000 signals generated in the same way.
Operating directly in the edge space, ${\bm L}_1$ and ${\bm L}_\mathrm{LG}$ are used as shift operators, once again using the absolute value of the flow signal when operating with ${\bm L}_\mathrm{LG}$.
Additionally, we consider an architecture that first determines the latent vector of node potentials ${\bm \Phi}={\mathcal D}({\bm A})\delta_v$, and then uses the graph Laplacian in the node space as an aggregation operator, similar to~\cite{GamaConvolutional}.
More precisely, we replace ${\bm f}$ by the estimated node potentials $\widehat{\bm \Phi}=({\bm B}^\top)^\dagger{\bm f}$ and ${\bm L}_1$ by ${\bm L_0}$ in~\eqref{eq:aggsample}.

The performance of these architectures on the test set is shown in Figure~\letterref{fig:exp}{C}.
Among the three approaches, the Hodge AGNN is superior in both accuracy and speed of convergence, confirming our intuition that it favorably models flow signals.

\section{Discussion}
\label{sec:disc}

In this work, we have considered the application of GNNs to flow signals supported on the edges of a graph.
To accomplish this, we use the Hodge Laplacian as a shift operator due to its favorable properties for modeling flows, which stem from the spectral decomposition into bases for cyclic and gradient flows.
We then demonstrated that this operator outperforms other graph shift operators in the tasks of flow interpolation and source localization.

%Although we focused on flow signals, the use of the Hodge Laplacian in a neural architecture opens up the possibility of applying these techniques to richer, higher-order data.
%That is, using the Hodge Laplacian as a descriptor of an abstract simplicial complex could allow for the processing of rich relational data, beyond that of a simple graph.
%One example of such an application is the network of drug-protein interactions~\cite{zitnik2018modeling}, where interactions do not always occur in a strictly pairwise fashion, and thus cannot be modeled completely by a simple graph.
%Rather, higher-order drug protein interactions are naturally represented by higher-order edges, which could then be understood through the lens of combinatorial Hodge theory combined with machine learning techniques.

Our ultimate goal is to develop a global and rigorous understanding of methods for processing and learning from data on \emph{higher-order networks}.
As a first step towards this objective, in this paper we kept a (non higher-order) graph as our domain, but considered the case where the data of interest is supported on the edges as opposed to the nodes of the graph.
The discussed use of the Hodge Laplacian in a neural architecture opens up the possibility of applying GNNs to data supported on arbitrarily high-order networks, allowing us to go beyond systems modeled by strictly pairwise interactions.

\bibliographystyle{IEEEtran}
\bibliography{ref}

\appendices

\section{Proof of Proposition~\ref{prop:sign-equivariance}}
\label{app:prop1}

Let ${\bm F}$ be an edge-flipping matrix where
\begin{equation}
  \label{eq:app:edgeflip}
  {\bm F}_{ij}
  \begin{cases}
    =0 & i\neq j \\
    \in\{-1,1\} & i=j
  \end{cases}.
\end{equation}
Applying this matrix to a flow signal ${\bm f}$ with associated Hodge Laplacian ${\bm L}_1$ yields ${\bm f}'={\bm F}{\bm f}$, ${\bm L}_1'={\bm F}{\bm L}_1{\bm F}$.
Substituting these into~\eqref{eq:rnnlayers1} yields
\begin{equation}
  \label{eq:app:inputflip}
  \begin{split}
    {\bm x}_0' &= {\bm f}' = {\bm F}{\bm f} = {\bm F}{\bm x}_0 \\
    {\bm x}_1' &= \frac{{\bm L}_1'}{\lambda_1}{\bm x}_0' = {\bm F}\frac{{\bm L}_1}{\lambda_1}{\bm F}{\bm x}_0' = {\bm F}\frac{{\bm L}_1}{\lambda_1}\bm{FF}{\bm x}_0 = {\bm F}{\bm x}_1 \\
    &\vdots \\
    {\bm x}_K' &= {\bm F}{\bm x}_K,
  \end{split}
\end{equation}
following from the fact that $\bm{FF}={\bm I}$.
That is, changing the edge orientation of the original flow signal changes the edge orientations of each element in the sequence of flow signals $\{{\bm x}_i\}_{i=0}^{K}$.

Applying a similar argument to~\eqref{eq:rnnlayers2}, we get that
\begin{equation}
  \label{eq:app:hiddenflip}
  {\bm H}_k' = \sigma\left({\bm F}\left({\bm x_k}{\bm u}^\top+{\bm H}_{k-1}{\bm V}\right)\right), 1\leq k\leq K.
\end{equation}
Clearly, if $\sigma(\cdot)$ is an odd function applied elementwise, \eqref{eq:app:hiddenflip} can be rewritten as
\begin{equation}
  \label{eq:app:hiddenflip-odd}
  {\bm H}_k'= {\bm F}\sigma\left({\bm x}_k{\bm u}^\top+{\bm H}_{k-1}{\bm V}\right) = {\bm F}{\bm H}_k, 1\leq k\leq K.
\end{equation}
Applying this to~\eqref{eq:rnnlayers3} then yields
\begin{equation}
  \label{eq:app:outputflip-odd}
  {\bm o}_K' = \sigma\left({\bm H}_K'{\bm w}\right) = \sigma\left({\bm F}{\bm H}_K{\bm w}\right) = {\bm F}\sigma\left({\bm H}_K{\bm w}\right) = {\bm F}{\bm o}_K,
\end{equation}
proving the ``if'' part of Proposition~\ref{prop:sign-equivariance}.
The ``only if'' part follows from a similar argument applied in reverse.

\section{Proof of Proposition~\ref{prop:agg-invariance}}
\label{app:prop2}

Again, let ${\bm F}$ be an edge-flipping matrix as described by~\eqref{eq:app:edgeflip}.
Additionally, let ${\bm C}$ be the row-selection matrix for the Hodge AGNN.

\noindent\textbf{Proof of i)} The first part of the Proposition~\ref{prop:agg-invariance} is true if the aggregated sequence ${\bm G}$ is equal to the sequence ${\bm G}'$ formed by the flipped flow signal.
So, for any flow signal ${\bm f}$ with associated Hodge Laplacian ${\bm L}_1$, as well as their flipped counterparts ${\bm f}'={\bm F}{\bm f}$, ${\bm L}_1'={\bm F}{\bm L}_1{\bm F}$, we have that
\begin{align}
  {\bm G}' &= {\bm C}\left[{\bm f}',\left(\frac{{\bm L}_1'}{\lambda_1}\right){\bm f}',\ldots,\left(\frac{{\bm L}_1'}{\lambda_1}\right){\bm f}'\right] \nonumber \\
  &= {\bm C}\left[{\bm F}{\bm f},{\bm F}\left(\frac{{\bm L}_1}{\lambda_1}\right)\bm{FF}{\bm f},\ldots,{\bm F}\left(\frac{{\bm L}_1}{\lambda_1}\right)\bm{FF}{\bm f}\right] \nonumber \\
  &= {\bm C}\left[{\bm F}{\bm f},{\bm F}\left(\frac{{\bm L}_1}{\lambda_1}\right){\bm f},\ldots,{\bm F}\left(\frac{{\bm L}_1}{\lambda_1}\right)^{E-1}{\bm f}\right] \nonumber \\
  &= {\bm C}{\bm F}{\bm C}^\top{\bm G}. \label{eq:app:aggflip1}
\end{align}
If ${\bm F}_{ii}=1$ for the edges selected by ${\bm C}$, then ${\bm C}{\bm F}{\bm C}^\top={\bm I}$.
Under these conditions, ${\bm G}'={\bm G}$, as desired.

\noindent\textbf{Proof of ii)} As shown previously, applying an edge-flipping matrix to a flow signal and its associated Hodge Laplacian in the AGNN architecture yields an aggregated signal ${\bm G}'=\bm{CFC}^\top{\bm G}$.
Notice that the matrix $\bm{CFC}^\top$ is itself an edge-flipping matrix, restricted to the edges selected by ${\bm C}$ (and thus, the edges measured in ${\bm G}$).

We split the parameters $\theta$ into $\{\theta_i\}_{i=1}^{\ell}$, where $\theta_j$ indicates the set of filters for the $j$th convolutional layer.
Consider the 1st layer of the CNN, with input ${\bm G}$ and output ${\bm Y}$.
The features (rows) of ${\bm Y}$ can be written as
\begin{equation}
  \label{eq:app:cnnfeats}
  {\bm Y}^j = \sigma\left(\sum_i{\bm G}^i\ast\theta_1^{ij}\right),
\end{equation}
where ${\bm Y}^j$ is the $j$th feature (row) of ${\bm Y}$, ${\bm G}^i$ is the $i$th feature (row) of ${\bm G}$, and $\theta_1^{ij}$ is a 1D convolutional filter.
Clearly, we can obtain the same output from ${\bm G}'$ by changing the sign of selected filters ${\theta_1}$, \ie
\begin{equation}
  \label{eq:app:cnnfeats-flipped}
  {\bm Y}^j = \sigma\left(\sum_{i}{\bm G}'^i\ast\left([\bm{CFC}^\top]_{ii}\theta_1^{ij}\right)\right).
\end{equation}
So, for a flow signal ${\bm f}$, Hodge Laplacian ${\bm L}_1$, and CNN parameters $\theta$, $\mathrm{AGG}_\theta({\bm f},{\bm L}_1)=\mathrm{AGG}_{\theta'}({\bm F}{\bm f},{\bm F}{\bm L}_1{\bm F})$, where $\theta_1'^{ij}=[\bm{CFC}^\top]_{ii}\theta_1^{ij}$, with the same parameters beyond the first layer, as desired.

Note that the first part of Proposition~\ref{prop:agg-invariance} is a special case of this, where $\bm{CFC}^\top={\bm I}\rightarrow\theta'=\theta$.

\end{document}